# SSMS - A Secure SMS Messaging Protocol for the M-Payment Systems [†]

Mohsen Toorani [‡]  Ali A. Beheshti

## Abstract

*The GSM network with the greatest worldwide number of users, succumbs to several security vulnerabilities. The Short Message Service (SMS) is one of its superior and well-tried services with a global availability in the GSM networks. The main contribution of this paper is to introduce a new secure application layer protocol, called SSMS, to efficiently embed the desired security attributes in the SMS messages to be used as a secure bearer in the m-payment systems. SSMS efficiently provides confidentiality, integrity, authentication and non-repudiation for the SMS messages. It provides an elliptic curve-based public key solution that uses public keys for the secret key establishment of a symmetric encryption. It also provides the attributes of public verification and forward secrecy. It efficiently makes the SMS messaging suitable for the m-payment applications where the security is the great concern.*

## 1. Introduction

The mobile communications has experienced a great acceptance among the human societies. At the beginning of 2007, the worldwide number of mobile users reached to 2.83 billion people [1]. The SMS is the most popular data bearer/service within GSM, IS-95, CDMA2000, and other cellular networks. It is a store-and-forward, easy to use, popular, and low cost service. While it is mainly used for the personal communications, it has also been used in applications where the other party is an information system. This includes a wide variety of applications ranging from remote control of the apparatus [2] to the m-banking and m-payment [3].

The *Global Service for Mobile communications* (GSM) with the greatest worldwide number of users suffers from many security problems [4]. In the GSM, only the airway traffic between the *Mobile Station* (MS) and the *Base Transceiver Station* (BTS) is optionally encrypted with a weak and broken stream cipher (A5/1 or A5/2). The authentication is unilateral and also vulnerable. The SMS messaging has some extra security vulnerabilities due to its store-and-forward feature, and the problem of fake SMS that can be conducted via the Internet. When a user is roaming, the SMS content passes through different networks and perhaps the Internet that exposes it to various vulnerabilities and attacks. Another concern is arisen when an adversary gets access to the phone and reads the previous unprotected messages. To exploit the popularity of SMS as a serious business bearer protocol, it is necessary to enhance its functionalities to offer the secured transaction capability. Data confidentiality, integrity, authentication, and non-repudiation are the most important security services in the security criteria that should be taken into account in many secure applications. However, such requirements are not provided by the traditional SMS messaging. Due the vast area of applications, each of them having a different level of security requirements, the most profitable solution is via the end-to-end security or the security at the application layer. It will be a network independent solution and does not need any change in the network's infrastructure. The security of SMS messaging at the application layer is considered in some literature [5-8]. The TS 03.48 standard [9] that was designed for the *SIM Application Toolkit* (SAT) can also be used for providing the end-to-end security in any SMS message outgoing or incoming to the *Subscriber Identity Module* (SIM).

In this paper, a new Secure SMS messaging protocol (SSMS) is introduced. Section 2 briefly describes the proposed protocol, section 3 takes a glimpse at its security attributes, and section 4 gives the conclusions.

---

[‡] Corresponding Author, ResearcherID: A-9528-2009

## 2. The Proposed Scheme (SSMS)

A brief description of our SSMS protocol is provided in this section. The end-to-end security in the cellular systems can be generally provided by exploiting the processing capabilities of one or some of the following items:
1) The *Mobile Equipment* (ME) using the programming languages,
2) The SIM card using SAT,
3) An additional smart card, e.g. JavaCard,
4) A crypto-processor that is embedded in the ME,
5) A portable PC (laptop) connected to the ME.

The SSMS provides the end-to-end security and is based on the first solution. It suggests using J2ME (Java 2 Mobile Edition) as the programming platform. J2ME is a runtime environment designed for resource restricted devices such as mobile phones and *Personal Digital Assistants* (PDA). It consists of *Connected Limited Device Configuration* (CLDC), and *Mobile Information Device Profile* (MIDP). A MIDlet is a program developed in J2ME. The MIDlet uses classes defined in APIs of CLDC and MIDP, and is run by the *Java virtual machine* (JVM). A J2ME application is a collection of byte code classes packed into a *Java Archive* file (JAR). However, it is also possible to use other platforms as they provide a way to access the extended feature of the SIM card. In J2ME, this feature is provided by the optional *Security and Trust Service API* (SATSA) package [10]. In the SATSA specifications, communications with the SIM card applications is defined via the APDU format of message exchange. In J2ME, the *Wireless Messaging API* (WMA) [11] provides the ability of sending and receiving SMS messages. The J2ME environment does not contain the cryptographic functions. However, the Lightweight API of the Bouncy Castle [12] can be used for this purpose. Obfuscators can also be used for both decreasing the JAR size and hardening the reverse-engineering of the produced codes. SSMS works on all Java enabled mobile phones supporting WMA and MIDP 2.0.

The SSMS protocol consists of the initialization phase, the message exchange phase in which the participants exchange their secured short messages, and the judge verification phase that is used when any dispute occurs. The revocation phase can follow the typical rules of *Public Key Infrastructures* (PKI) and will not be considered in this paper.

### 2.1. Initialization

The initialization phase of the SSMS includes:
1) Selecting the domain parameters.
2) Registering the user details into the system, generating the public/private keys, and issuing a certificate for the public key of each user.
3) Installing the application software on the mobile phone.

The domain parameters consist of a suitably chosen elliptic curve $E$ defined over the finite field $F_q$ with the Weierstrass equation of the form $y^2 = x^3 + ax + b$ and a base point $G \in E(F_q)$ in which $q$ is a large prime number. In order to make the elliptic curve non-singular, $a, b \in F_q$ should satisfy $4a^3 + 27b^2 \neq 0 (\mod q)$. To guard against small subgroup attacks [13], the point $G$ should be of prime order $n$ ($nG = O$ where $O$ denotes the point of elliptic curve at infinity), and we should have $n > 4\sqrt{q}$. To guarantee the intractability of ECDLP to the Pollard-rho and Pohlig-Hellman algorithms [14], $n > 2^{160}$ is recommended. To protect against other known attacks on special classes of elliptic curves, $n$ should not divide $q^i - 1$ for all $1 \leq i \leq V$ ($V = 20$ suffices in practice [15]), $n \neq q$ should be satisfied, and the curve should be non-supersingular [13].

A strong block cipher such as AES is recommended to be chosen for encrypting the messages. It will be used in the core of SSMS protocol. It is also feasible to provide a list of block ciphers and let the user choose the suitable algorithm. However, it is not recommended since it can increase the code size of the application.

The private key of user $U$ is a randomly selected integer $SK_U \in_R [1, n-1]$, and the corresponding public key is generated as $PK_U = SK_U G$. Two different approaches can be followed for the public/private key generation:
1) Generating the public/private keys in a *Key Generating Server* (KGS), as is depicted in Figure 1.
2) Generating the public/private keys in the ME, as is depicted in Figure 2.

In the first approach that is recommended by the SSMS, the public/private keys are directly and securely stored on the SIM cards. The mobile phones are not capable of generating very strong random numbers to be used as strong private keys so the first approach is preferred as it provides more security. The second approach, on the other hand, has more flexibility and may be suitable for some applications. However, several precautions should be taken into account when using the second approach: The *Certificate Authority* (CA) should verify that each entity really possesses the corresponding private key of its claimed public key. This may be accomplished by a zero-knowledge technique. To thwart the *invalid-curve* attacks, the CA server should also check the validity of public keys [13].

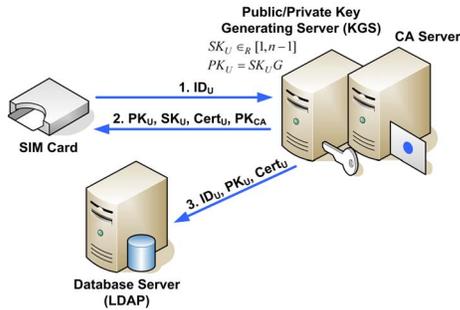

**Figure 1. When the key generation is taken place in a KGS**

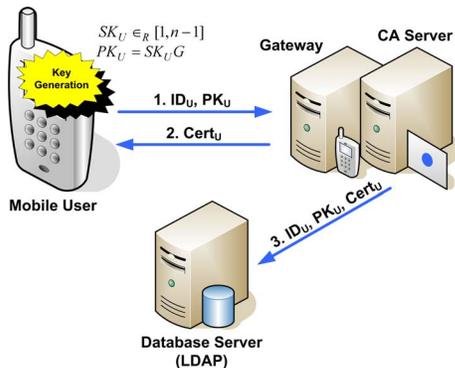

**Figure 2. When the key generation is taken place in the ME**

The CA server issues a certificate $Cert_U$ for the public key of each user. The certificates contain strings of information that uniquely identify users and bind their identities to their public keys. The X.509v3 certificates are a popular type of certificates that are extensively used. Each user of the GSM network has several unique identities such as IMSI, MSISDN, and etc. For making the SSMS convenient, the unique international phone number of each user (MSISDN) is chosen as his/her identifier ($ID_U$). The unique identifier, the public key, and the corresponding certificate of each user are also stored in an LDAP (Lightweight Directory Access Protocol) directory, as is depicted in Figures 1 and 2.

It is assumed that the participants have access to an authentic copy of the *CA*'s public key, in order to use it for the certificate validation. The private key is encrypted and stored in an elementary file of the SIM Card. The private key can be encrypted using a user's password or PIN (Personal Identification Number). The public keys can be stored in a transparent elementary file. According to the GSM 11.11 standard, the trusted keys and certificates (e.g. CA's) are stored in 4FXX files. The key information can be stored in file 4F50. The hash value of the private key that can be used for validation of the encrypting password can also be stored in the SIM Card.

The application can be purchased and be manually installed on the mobile phone if the mobile phone is capable of supporting such an application. It is also feasible to download and install the application via an *Over-The-Air* (OTA) server, as is depicted in Figure 3. The *Application Management Software* (AMS) takes care of downloading and installing the application on the device. It is also feasible to combine the application installation and key generation processes if the key generation is to be done in the mobile phone [8].

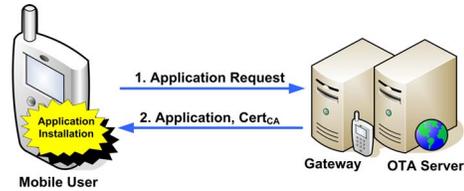

**Figure 3. Application installation via an OTA Server**

### 2.2. The Message Exchange

*Alice* as the sender or initiator of the protocol wants to securely send her message *M* to *Bob*. Her message can be a payment order and *Bob* may be a bank or a service provider. SSMS is a certificate-based protocol so it is necessary for both of participants to perform the certificate validation [14] for the certificate of the other party and use it for verifying the corresponding public key. Since the *Certificate Revocation List* (CRL) is too large for the limited memory capacity of mobile devices, SSMS uses the *Online Certificate Status Protocol* (OCSP) [16] for checking the revocation status of the certificates. It has another advantage due to the online and timely inquiry feature of the OCSP. However, the OCSP server's duties in the SSMS differ from what is specified in RFC 2560. The OCSP server in the SSMS should also verify the public key of the corresponding user if the queried certificate has a "good" status. The result of such verification should be included in the OCSP responses. The OCSP responses are digitally signed with a private key that its corresponding trusted public key is known to the participants. Basic configuration of the SSMS protocol is depicted in Figure 4 in which $OCSP_A$ and $OCSP_B$ are the OCSP tokens for the certificates of *Alice* and *Bob*. The message exchange for the basic SSMS configuration is as follows.

**SSMS Composing:** *Alice* queries the OCSP server via the *Bob*'s phone number ($ID_B$) for his public key and the revocation status of his certificate by sending an SMS to the gateway of the OCSP server. The OCSP server produces an OCSP response, digitally signs it, and sends ($ID_B$, $PK_B$, $OCSP_B$) to *Alice* perhaps through SMS. *Alice* verifies the OCSP server's signature and uses $ID_B$ and

$PK_B$ to generate $(R,C,s)$ from the message $M$ by following the below steps:

(1) Randomly selects an integer $r \in_R [1, n-1]$.
(2) Computes $R = rG = (x_R, y_R)$.
(3) Computes $K = (r + \tilde{x}_R SK_A)PK_B = (x_K, y_K)$ where $\tilde{x}_R = 2^{\lceil f/2 \rceil} + (x_R \mod 2^{\lceil f/2 \rceil})$ in which $f = \lfloor \log_2 n \rfloor + 1$. If $K = O$ she goes back to step (1). Otherwise, she drives the secret key of encryption as $k = H'(x_K \| ID_A \| y_K \| ID_B)$ in which $H'$ is a one-way hash function that generates the required number of bits for the secret key of the deployed symmetric encryption ($l$ bits).
(4) Computes the ciphertext as $C = E_k(M)$ where $E(.)$ denotes the approved block cipher.
(5) Computes $s = tSK_A - r (\mod n)$ where $t = H(M \| x_R \| ID_A \| y_R \| ID_B \| k)$.
(6) Sends $(R,C,s)$ to *Bob* through SMS.

The *Short Message Center* (SMC) of the network operator automatically adds timestamps to the exchanged short messages. A port number is also assigned to the message indicating that the message is intended for the SSMS application so the AMS will automatically start the corresponding MIDlet upon receiving the message.

Each short message can contain 140×8 bits of user data. The GSM 03.38 standard [17] increases the maximum length of a text-based SMS to 160 (7-bits) characters. For the longer messages, new features such as SMS concatenation and SMS compression have been added in phase 2+ and are supported by the newer mobile phones. It is possible to concatenate several short messages to send messages or data records that are longer than 140 bytes. Each segment of the short message is still limited to 140 bytes but the receiving application is alerted that the entire message is not contained within one segment. The network handles segments of a concatenated message like any other short message and the relationship between the segments is only made at the end-entities. Although it is theoretically possible to concatenate 255 short messages, it is actually limited to five or so [18].

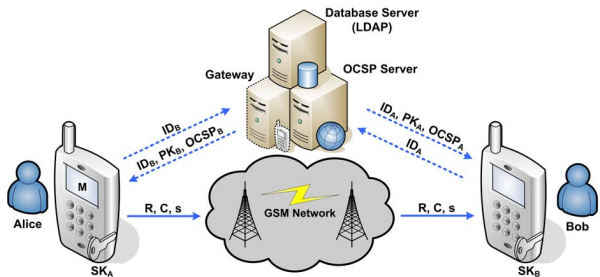

**Figure 4. Basic configuration of the SSMS**

**SSMS Delivering:** Upon receiving an SSMS containing $(R,C,s)$, the AMS on the *Bob* side automatically starts the SSMS application. The application may check whether the timestamp resides in an acceptable time window. It queries the OCSP server via the *Alice*'s phone number ($ID_A$) for her public key and the revocation status of her certificate by sending an SMS to the gateway of the OCSP server. The OCSP server produces an OCSP response, digitally signs it, and sends ($ID_A$, $PK_A$, $OCSP_A$) to *Bob* perhaps through SMS. The SSMS application on the *Bob* side verifies the OCSP server's signature and follows the following steps to extract the *Alice*'s message and verify her signature.

(1) Checks the validity of $R = (x_R, y_R)$ by verifying that all the following conditions are satisfied. Otherwise, he terminates the process with failure.
 (1.1) $R \neq O$.
 (1.2) $x_R$ and $y_R$ should have the proper format of $F_q$ elements.
 (1.3) $R$ should satisfy the defining equation of $E$.
(2) Computes $K = SK_B(R + \tilde{x}_R PK_A) = (x_K, y_K)$ and derives the secret key of encryption as $k = H'(x_K \| ID_A \| y_K \| ID_B)$.
(3) Decrypts the received ciphertext as $M = D_k(C)$.
(4) Computes $t = H(M \| x_R \| ID_A \| y_R \| ID_B \| k)$.
(5) Verifies whether $sG + R = tPK_A$. If this is true, he accepts $M$ as the correct plaintext that is sent by *Alice* and sends her a confirmation message $M'$ perhaps in addition to a tag $t' = MAC_k(M')$. Otherwise, he rejects $M$ and reports it as a malicious message to the service provider.

Due to the processing and memory constraints, delegating validations to a trusted server will offer significant advantages. The required time of sending a certificate to a validation server, receiving, and authenticating the response can be considerably less than the required time of performing the certificate path discovery and validation. The SSMS performance can be greatly improved by delegating the validation process to a *Delegated Validation* (DV) server, as is depicted in Figure 5. The DV server is completely independent of the wireless network, e.g. GSM. It can be independently provided by the m-payment service provider. The DV server accomplishes the certificate validation via the *Delegated Path Validation* (DPV) protocol [19]. However, its duties defer from what is specified in RFC 3379. All the short messages will be directed to the DV server through the GSM network. The DV server accomplishes the certificate and public key validations for both sender and recipient. It queries the database server for the certificates of participants via their MSISDN. It obtains the revocation statuses by getting

OCSP responses from the OCSP server. It also checks the validity of point *R*. The DV server will contact the designated recipient (via SMS) after a successful validation. If any error occurs, the DV server sends an error message to the initiator and saves a copy in its log file. According to the security policies, all the transmitted messages may be separately saved by the DV server for the possible disputes. The DV server digitally signs its responses unless an error is occurred. The signed responses should include a hash value of all the transmitted parameters in addition to the identifiers of both sender and recipient.

In the optimized SSMS configuration, the first step of the basic SSMS delivering phase is omitted. *Bob* does not check the validity of point *R* because it is checked by the DV server. He just checks the DV server's signature. *Alice* does not need to get any OCSP response for the *Bob*'s certificate. She just needs to know the public key of *Bob* and save it in his phone for the future uses. If she does not know the public key of *Bob*, she may query it from the database server, as is depicted in Figure 5. Therefore, the computational costs and communication overheads are efficiently decreased by the optimized SSMS configuration.

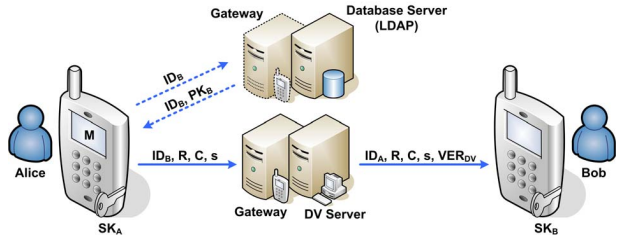

**Figure 5. Optimized configuration of the SSMS for the Customer-to-Customer (C2C) model**

## 2.3. The Judge Verification in disputes

When *Bob* claims that he has received $(R,C,s)$ from *Alice*, the trusted third party (judge) wants him to provide $(R,C,M,k,s)$. *Bob* simply extracts *M* and *k* from the previously saved $(R,C,s)$. The judge follows the following steps to decide on what *Bob* claims.

(1) Checks the validity of $Cert_A$ and uses it for verifying $PK_A$.
(2) Verifies whether $M = D_k(C)$. If this is not the case, *Bob* is wrong.
(3) Computes $t = H(M \| x_R \| ID_A \| y_R \| ID_B \| k)$.
(4) Verifies the *Alice*'s signature by checking the $sG + R = tPK_A$ condition. If this is not the case, *Bob* is wrong. Otherwise, *Alice* has sent $(R,C,s)$ to *Bob*.

If the DV server saves the transmitted messages, the judge as an additional proof may query the DV server to confirm that *Alice* has sent a message containing $(R,C,s)$ to *Bob*.

## 3. The Security of SSMS

The correctness of SSMS can be simply verified. *Alice* and *Bob* reach to the same secret key of symmetric encryption as:

$$K_A = (r + \widetilde{x}_R SK_A)PK_B = (r + \widetilde{x}_R SK_A)SK_B G =$$
$$= SK_B(rG + \widetilde{x}_R SK_A G) = SK_B(R + \widetilde{x}_R PK_A) = K_B$$

The session key derivation function of the SSMS is an improved version of the HMQV key establishment protocol [20]. However, it does not exactly correspond with the HMQV specifications. Defining $\widetilde{x}_R$ as the least significant half in binary representation of $x_R$ is just a trade-off between security and efficiency. The validation of *R* in the SSMS delivering phase is accomplished to prevent the *invalid-curve* attack. Hereunder, some security attributes of the SSMS are briefly described.

1) *Confidentiality*: The confidentiality is completely resided in the secrecy of session key since SSMS uses a strong block cipher. The session key differs for different sessions and is derived from the private keys of the participants. The *Unknown Key-Share* (UKS) attack is thwarted because the *Alice*'s identifier is involved in derivation of session key *k*. There are only two ways to defeat the confidentiality: finding $SK_B$, or having both of $SK_A$ and *r*. Deducing the corresponding *r* of *R* is generally in deposit of solving the *Elliptic Curve Discrete Logarithm Problem* (ECDLP) that is computationally infeasible with the chosen domain parameters.

2) *Authentication*: The implicit authentication is provided in three ways: both the participants verify the certificate of the other party; only the legitimate parties with the true private keys can reach to a correct key agreement; and whenever the recipient verifies the signature.

3) *Unforgeability*: It is computationally infeasible to forge the signature of *Alice* without having her private key $SK_A$.

4) *Integrity*: The hash value of message, concatenated with some variable parameters is involved in the signature generation. The integrity is guaranteed by the security attributes of the deployed hash function and also the unforgeability of the signature.

5) *Non-repudiation*: It can be easily deduced from the unforgeability.

6) *Forward secrecy of message confidentiality*: As a one-pass protocol, we cannot prospect SSMS for the *Perfect Forward Secrecy* (PFS). However, if $SK_A$ is revealed, the attacker cannot recover the previous messages since he should know the respective random number *r* of the corresponding session that is generally in deposit of solving the ECDLP.

7) *Public verifiability*: Given $(R,C,M,k,s)$, anybody can verify the signature without any need for the private keys of the participants.

# 4. Conclusions

A new public key-based solution for secure SMS messaging (SSMS) is introduced in this paper. It is an application layer protocol that simultaneously provides the confidentiality, integrity, authentication, non-repudiation, public verification, and the forward secrecy of message confidentiality. It efficiently combines encryption and digital signature and uses public keys for a secure key establishment to be used for encrypting the short messages via a symmetric encryption. Since it deploys elliptic curves and a symmetric encryption algorithm, it has great computational advantages over the previously proposed public key solutions while simultaneously providing the most feasible security services. It has great advantages to be used in the real m-payment applications and whenever the secure SMS messaging is important. However, its structure does not reply on the SMS messaging and is suitable for any other store-and-forward technology.